\documentclass{sig-alternate}

\usepackage{verbatim}

\begin{document}

\title{Dealing with Large Schema Sets in \\ Mobile SOS-Based Applications}
\numberofauthors{3}

\author{
\alignauthor
Alain Tamayo\\
       \affaddr{Institute of New Imaging Technologies}\\
       \affaddr{Universitat Jaume I, Spain}\\
       \affaddr{Ave Vicent Sos Baynat, SN, 12071, Castell\'on de la Plana}\\
       \email{atamayo@uji.es}
\alignauthor
Carlos Granell\\
      \affaddr{Institute of New Imaging Technologies}\\
       \affaddr{Universitat Jaume I, Spain}\\
       \affaddr{Ave Vicent Sos Baynat, SN, 12071, Castell\'on de la Plana}\\
       \email{carlos.granell@uji.es}
\alignauthor Joaqu\'in Huerta\\
      \affaddr{Institute of New Imaging Technologies}\\
       \affaddr{Universitat Jaume I, Spain}\\
       \affaddr{Ave Vicent Sos Baynat, SN, 12071, Castell\'on de la Plana}\\
       \email{huerta@uji.es}
}

\maketitle
\begin{abstract}
Although the adoption of OGC Web Services for server, desktop and web applications has been successful, its penetration in mobile devices has been slow. One of the main reasons is the performance problems associated with XML processing as it consumes a lot of memory and processing time, which are scarce resources in a mobile device. In this paper we propose an algorithm to generate efficient code for XML data binding for mobile SOS-based applications. The algorithm take advantage of the fact that individual implementations use only some portions of the standards' schemas, which allows the simplification of large XML schema sets in an application-specific manner by using a subset of XML instance files conforming to these schemas.
\end{abstract}




\section{Introduction} 

\textit{Interoperability} is a key concept when building distributed applications, as it ensures that service providers and consumers can exchange information in a way that can be understood. In the Geographic Information Systems (GIS) field, this interoperability is achieved by using standards or \textit{implementation specifications}, such as those defined by the Open Geospatial Consortium (OGS), known as \textit{OGC Web Services } (OWS). These standards allow clients to access geospatial data through a well-defined set of operations. The specifications define the structure of XML messages exchanged between clients and servers using XML Schema \cite{w3c:schemas1, w3c:schemas2}.

One of these standards is the Sensor Observation Service (SOS) Implementation Specification \cite{ogc:sos}, which allows the publication and consumption of information gathered by sensors or system of sensors. This specification has gained a lot of popularity in recent years, apparently because of the explosion of the number of sensors and related devices producing a massive amount of data \cite{article:economist}. Several implementations of this specification have been presented for the client and server side mainly targeted to servers, desktop and web applications. As the adoption of the standard for these applications have been successful, its integration in mobile devices has been slow. One of the main reasons is the performance problems associated with XML processing, as the effort to parse and serialize XML messages from files (or communication channels) to memory and vice versa, consumes a lot of memory and processing time, which are scarce resources in a mobile device. 

According to \cite{proc:white} XML processing can be implemented using a \textit{vocabulary-independent data access interface} such as those provided by SAX\footnote{http://www.saxproject.org/}  or DOM\cite{w3c:dom}; or using a \textit{vocabulary-dependent data access interface}, where XML data is mapped into application-specific concepts. The first option is recognized to be difficult and error-prone producing code that is hard to modify and maintain. The second option, also known as XML Data Binding, is favoured as \textit{`` relieve developers from the burden of mapping data from a vocabulary-independent DAI (Data Access Interface) to application- specific data structures. Developers can focus on the semantics of the data they are manipulating, while leaving the type conversion to the vocabulary-specific DAI implementation''}  \cite{proc:white}. XML data binding code is often produced by using code generators. Code generators provide an attractive approach, potentially giving benefits such as increased productivity, consistent quality throughout all the generated code, and the potential to support different programming languages, frameworks and platforms.

Recent studies have proven that XML can be processed efficiently in resource-constrained devices if the appropriate methods and tools are used \cite{article:kangasharju1, proc:lindholm}. There are also several tools available for generating XML data binding code for mobile devices such as XBinder\footnote{http://www.obj-sys.com/xbinder.shtml} and CodeSysnthesis eXSD\footnote{http://www.codesynthesis.com/products/xsde}, or for building complete web services communication end-points for resource constrained environments, such as gSOAP \cite{proc:vanengelen}. Nevertheless, these solutions are not easily or effectively applied to mobile SOS-based applications. For example, the solutions presented in \cite{proc:kabisch, proc:kangasharju2, w3c:exi} for efficient processing of XML in mobile devices use compression techniques for XML data. This requires that all of the existing infrastructure of data providers must be modified to offer data in these compressed formats.  On the other hand, XML data binding code generators tend to map types in schema files to types in the target language in a straightforward way by creating a type in the target programming language for every type in the schema files, which cause that large schema files produces large binary compiled files. Last, several OWS specification do not provide support or provide a limited amount of support for SOAP. Requests sent in XML format over HTTP are widespread, although the current trend is to support SOAP in new specifications \cite{ogc:httpSoap}.

The main hurdle to generate efficient code for mobile applications in SOS is the large size and complexity of the schemas associated to the specification. The large size of the schemas is justified because they must satisfy a large set of usage scenarios, although individual implementations frequently use only a small fraction of them. This fact offers a way of optimizing real implementations by using only the subset of the schemas really necessary for a given application. Based on this, we present in this paper, an algorithm to simplify large XML schema sets, such the one associated to SOS, in an application-specific manner by using a set of XML instance files conforming to these schemas. A real use case scenario, the implementation of a  mobile SOS client for the Android platform, is presented to prove the effectiveness of the algorithm.

The remainder of this paper is structured as follows. Next section presents an introduction to XML Schema, including necessary notation and concepts used in this paper. Section 3 presents the algorithm to simplify schema sets based in a subset of input instance files. Section 4 presents experimental results using an use case scenario. In Section 5 related work on the subject is presented. Lastly, we present conclusions and future work.

\section{XML Schema}

XML Schema is used to define the structure of information contained in XML instance files \cite{w3c:schemas1, w3c:schemas2}. The structure is defined using schema components such as complex types, simple types, elements, attributes, and element and attribute groups. An instance document conforming to this structure is said to be valid against the schema. We denote the set of all valid files against a schema \textit{S} as \textit{I(S)}. Figure \ref{fig1} shows a fragment of a schema file. The file contains the declaration of three global complex types and a global element. For the sake of simplicity we have omitted the schema root element and namespace declarations.

In Figure \ref{fig2} we can see two valid instance documents for this schema. In the second instance we can observe that the \textit{item} element is of type \textit{Child} instead of type \textit{Base}. This is because XML Schema provides a derivation mechanism to express subtyping relationships. This mechanism allows types to be defined as subtypes of existing types, either by extending the base types content model in the case of \textit{derivation by extension} (\textit{Child} in Figure \ref{fig1}); or by restricting it, in the case of \textit{derivation by restriction}. What is interesting about type derivation is that wherever we find in the schemas an element of type A, the actual type of the element in an instance file can be either A or any type derived from A. This is why in the example an element of type \textit{Base} can be substituted by an element of type \textit{Child}. This polymorphic situation creates non-explicit dependencies between types, which we call \textit{hidden dependencies}.

\begin{figure}[!h]
\raggedright
<complexType name=``Base''>\\
\hspace{3mm}<sequence>\\
\hspace{6mm}<element name=``baseElem''  type=``string''/>\\
\hspace{6mm}<element ref=``baseElem2''  minOccurs=``0''/>\\
\hspace{3mm}</sequence>\\
</complexType>\\ 
\vspace{5mm}
<complexType name=``Child''>\\
\hspace{3mm}<complexContent>\\
\hspace{6mm}<extension base=``Base''>\\
\hspace{9mm}<sequence>\\
\hspace{12mm}<element name=``chdElem'' \\
\hspace{15mm} type=``string''/>\\
\hspace{9mm}</sequence>\\
\hspace{6mm}</extension>\\
\hspace{3mm}</complexContent>\\
</complexType>\\ 
\vspace{5mm}
<complexType name=``ContainerType''\\
\hspace{3mm}<sequence>\\
\hspace{6mm}<element name=``item'' type=``Base''\\
\hspace{9mm}maxOcurrs=``unbounded''/>\\
\hspace{3mm}</sequence>\\
</complexType>\\

<element name=``container'' type=``ContainerType'' /> \\ 
\vspace{5mm}
<element name=``baseElem2'' type=``string'' /> \\ 
\caption{XML Schema file fragment.}\label{fig1}
\end{figure}

\begin{figure}[!h]
\raggedright
\textbf{Instance 1:}\\
<Container>\\
\hspace{3mm}<item>\\
\hspace{6mm}<baseElem> String Value 1</baseElem>\\
\hspace{3mm}</item>\\
\hspace{3mm}<item>\\
\hspace{6mm}<baseElem> String Value 2</baseElem>\\
\hspace{3mm}</item>\\
</Container>\\
\vspace{5mm}
\textbf{Instance 2:}\\
<Container>\\
\hspace{3mm}<item xsi:type="Child">\\
\hspace{6mm}<baseElem> Base String Value</baseElem>\\
\hspace{6mm}<chdElem> Child String Value</chdElem>\\
\hspace{3mm}</item>\\
</Container>\\

\caption{Valid XML fragments for schema fragment in Figure 1.}\label{fig2}
\end{figure}

Apart from type derivation, a second subtyping mechanism is provided through \textit{substitution groups}. This feature allows global elements to be substituted by other elements in instance files. A global element E, referred to as \textit{head element}, can be substituted by any other global element that is defined to belong to the E's substitution group. 

In the following subsection we introduce the notation used in the remainder of the report to refer to nodes and components included into XML instance documents and schema files, respectively. We also define concepts, relations and operations necessary to expose our algorithm. After this we present a brief description of SOS schemas.

\subsection{Notation}

To refer to nodes contained in instance files, we will use \textit{XPath} notation \cite{w3c:xpath}. XPath expressions are shown in bold. The following examples referring to nodes in Figure 2 should suffice to understand the notation through the remainder of this paper:
\begin{itemize}
\item \textbf{/Container} refers to the root node of instance files.
\item \textbf{/Container/item} refers to all items contained in the root elements.
\item  \textbf{/Container/item[i]} refers to item in position \textit{i} inside \textbf{/Container}. Positions are counted starting at 1.
\end{itemize}
To refer to components in schema files, we will use to following notation:
\begin{itemize}
\item To refer to global types and elements, we use its name in italics, e.g. \textit{Container}, \textit{ContainerType}, etc.
\item To refer to attributes or elements within types, model groups or attribute groups, we add their name and a colon as prefix to the attribute or element name. The whole expression is written in italics. For example: \textit{ContainerType:item}, \textit{Base:baseElem}, \textit{Child:chdElem}, etc.
\end{itemize}
For the purpose of our discussion we define the concept of schema set in the following way:\\

\textit{ \textbf{Definition 1}: An schema set $ S = (T_{S}$, $ E_{S}$, $A_{S}$, $MG_ {S}$, $AG_{S}$, $R_{S})$, where $T_{S}$ is the set of all type definitions, $ E_{S}$ is the set of all element declarations, $A_{S}$ is the set of all attribute declarations, $MG_{S}$ is the set of all element group definitions, $AG_{S}$ is the set of all attribute group definitions, and $R_{S}$ is a set of binary relations (described later) between components of $T_{S}$, $E_{S}$, $A_{S}$, $MG_{S}$, and $AG_{S}$. }\\

Components included in sets $T_{S}$, $MG_{S}$ and $AG_{S}$, are composed by a set of inner components. In the case of types, inner components can refer to global elements, attributes, model groups and attribute groups, or they can be nested element and attribute declarations. Model groups may contain references to global elements and other model groups, or they may contain nested element declarations. Similarly, attribute groups may contain references to other global attributes and attribute groups, or they may contain nested attribute declarations. Inner components can be \textit{optional}, meaning that is legal that they do not appear in all valid instance documents. For example, element \textit{baseElem2} in \textit{Base} is optional; as such, items in Figure \ref{fig2} are valid even if they do not contain this element. 

The binary relations contained in $R_{S}$ are:
\begin{itemize}
\begin{sloppypar}
\item \textit{\textbf{isOfType(x, t)}}: relates an element or attribute \textit{x} to its corresponding type \textit{t}. For example: \textit{\textbf{isOfType}}(\textit{Container}, \textit{ContainerType}), \textit{\textbf{isOfType}}(\textit{Base:baseElem},  \textit{string}).
\end{sloppypar}
\item \textit{\textbf{reference(x, y)}}: relates $x \in T_{S} \cup MG_{S} \cup AG_{S}$ to $y \in E_{S}  \cup A_{S} \cup MG_{S} \cup AG_{S} $ if \textit{x} references \textit{y} in its definition using the \textit{ref} attribute in any of its components, e.g. \textit{\textbf{reference}}(\textit{Base, baseElem2}).
\item \textit{\textbf{contains(x, y)}}: relates $x \in T_{S} \cup MG_{S} \cup AG_{S}$ to $y \in E_{S} \cup A_{S}$ if \textit{x} defines \textit{y} as an inner attribute or element in its declaration, e.g. \textbf{\textit{contains}}(\textit{Base},\textit{Base:baseElem}), \textbf{\textit{contains}}(\textit{Child},\textit{child:chdElem}), \textbf{\textit{contains}}(\textit{Container},\textit{ Container:item}).
\item \textit{\textbf{isDerivedFrom(t, b)}}: relates a type \textit{t} to its base type \textit{b}, e.g. \textit{\textbf{isDerivedFrom}}(\textit{Child}, \textit{Base})  
\item \textit{\textbf{isInSubstitutionGroup(x, y)}}: relates an element \textit{x} to another element y if \textit{y} is the head element of the x's substitution group. 
\end{itemize}

The schema set S for the schema fragment in Figure \ref{fig1} remains as follows\footnote{XML Schema \textit{anyType} has been omitted purposely to simplify exposition.}:

\textit{
\begin{tabbing}
 S  = \{ \= $T_{S}$ = \{ Base, Child, string, ContainerType\}\\
\>  $ E_{S}$ = \{ \= Container, baseElem2, Base:baseElem,  \\
\> \> Child:chdElem, ContainerType:item\} \\
\> $ A_{S}$ = $\emptyset$ \\ 
\> $MG_{S} $= $\emptyset $\\
\>  $AG_{S} $= $\emptyset$  \\
\>  $R_{S} $  = \{ \= isOfType =  \{ \= (Container,ContainerType), \\
\> \> \>(baseElem2, string),  \\
\> \> \>(Base:baseElem, string), \\ 
\> \> \>(Child:chdElem, string), \\
\> \> \>(ContainerType:item, Base) \}, \\
\> \> isDerivedFrom = \{(Child, Base)\},\\
\> \> reference = \{(Base, baseElem2)\},\\
\> \> contains = \{\= (Base, Base:baseElem),\\
\> \> \>(Child, Child:chdElem),\\
\> \> \>(Con\=tainerType,\\
\> \> \> \> ContainerType:item)\}\\
\> \> isInSubstitutionGroup = $\emptyset$\}\\
\end{tabbing}
}

\begin{figure}[!h]
  \centering
  \includegraphics[scale=0.6]{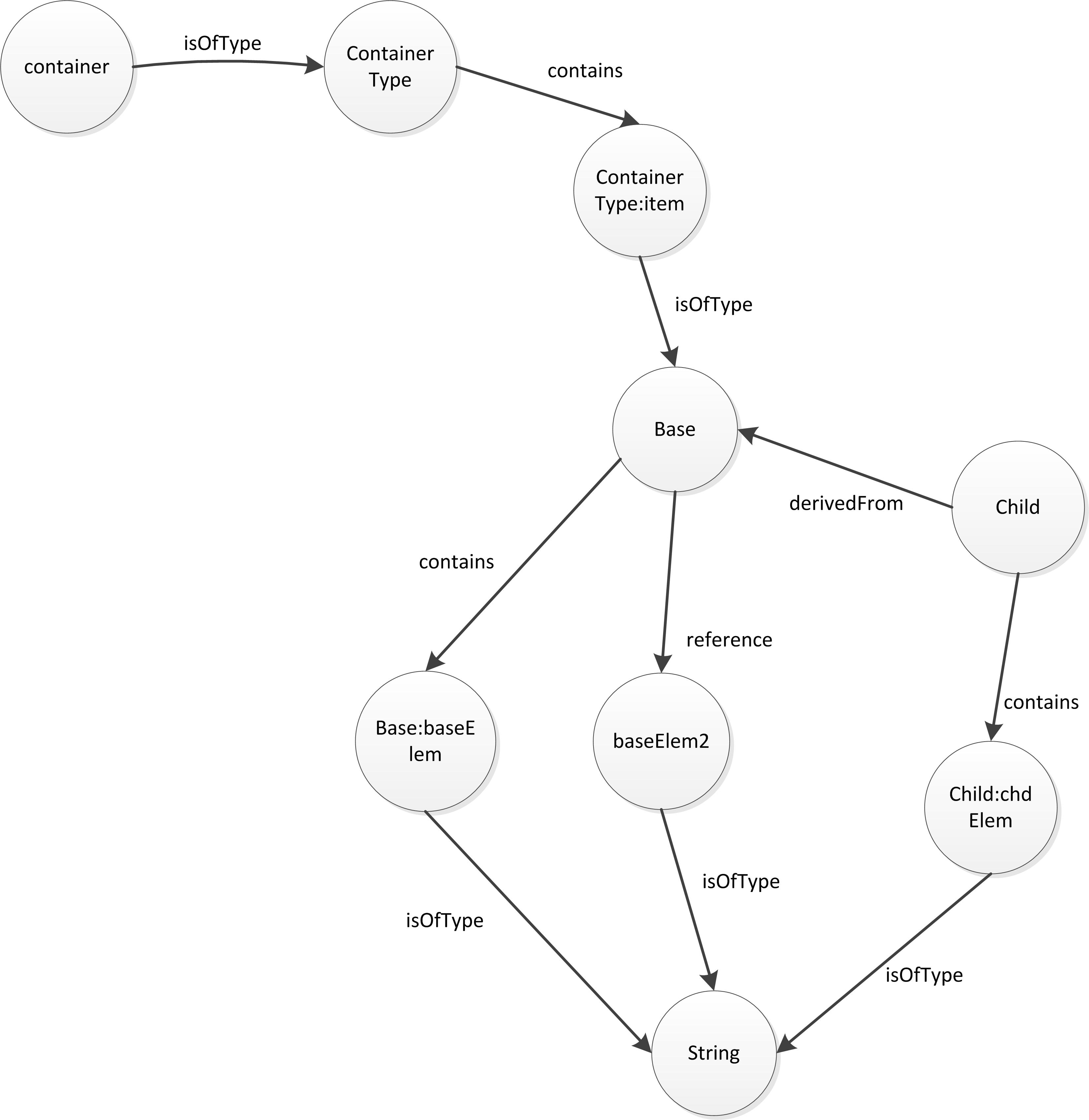}\\
  \vspace{5mm}
  \caption{Graph of relations in schema fragment in Figure 1}\label{fig4}
\end{figure} 

Figure \ref{fig4} shows a graph with all of these relations. This graph does not reflect hidden dependencies between types or elements. To include them in the graph we had to add an extra edge between \textit{ContainerType:item} and \textit{Child} as the former may be of type \textit{Child} in instance files.

Next, we define the \textit{subset} relation for schemas:\\

\textit{\textbf{Definition 2}: Let S = ($T_{S}$, $E_{S}$, $A_{S}$, $MG_{S}$, $AG_{S}$, $R_{S}$) and $S_{1}$=($T_{S1}$, $E_{S1}$, $A_{S1}$, $MG_{S1}$, $AG_{S1}$, $R_{S1}$),  be two schema sets, we said that $S_{1}$ is a subset of  \textit{S} if  $T_{S1} \subseteq T_{S}$,  $E_{S1} \subseteq E_{S}, A_{S1} \subseteq A_{S}, MG_{S1} \subseteq MG_{S}, AG_{S1} \subseteq AG_{S}$, and for every relation $R_{iS}$ in $R_{S}$ , $R_{iS1} \subseteq R_{iS}$, for example, $\textbf{isTypeOf}_{RS1}$ $\subseteq  \textbf{isTypeOf}_{RS}$ \\}

According to this definition a subset of the schema set in Figure \ref{fig1} could be:

\textit{
\begin{tabbing}
 S  = \{ \= $T_{S}$ = { Base, string, ContainerType}\\
\>  $ E_{S}$ = \{\= Container, Base:baseElem,  \\
\> \> ContainerType:item\} \\
\> $ A_{S}$ = $\emptyset$ \\ 
\> $MG_{S} $= $\emptyset $\\
\>  $AG_{S} $= $\emptyset$  \\
\>  $R_{S} $  = \{ \= isOfType =  \{ \= (Container,ContainerType), \\
\> \> \>(Base:baseElem, string), \\ 
\> \> \>(ContainerType:item, Base) \}, \\
\> \> isDerivedFrom = $\emptyset$,\\
\> \> reference = $\emptyset$,\\
\> \> contains = \{\= (Base, Base:baseElem),\\
\> \> \>(Con\=tainerType,\\
\> \> \> \> ContainerType:item)\}\\
\> \> isInSubstitutionGroup = $\emptyset$\}\\
\end{tabbing}
}

Last, we define the \textit{union} of two schema sets. This operation will be used in the following sections.\\

\textit{\textbf{Definition 3}: Let $S_{1}$ = ($T_{S1}$, $E_{S1}$, $A_{S1}$, $MG_{S1}$, $AG_{S1}$, $R_{S1}$) and $S_{2}$=($T_{S2}$, $E_{S2}$, $A_{S2}$, $MG_{S2}$, $AG_{S2}$, $R_{S2}$),  be two schema sets, we said that S = ($T_{S}$, $E_{S}$, $A_{S}$, $MG_{S}$, $AG_{S}$, $R_{S}$)  is the union of $S_{1}$ and $S_{2}$ if  $T_{S}$ = $T_{S1} \cup T_{S2}$, $E_{S}$ = $E_{S1} \cup E_{S2}$, $MG_{S}$ = $MG_{S1} \cup MG_{S2}$, $A_{S}$ = $A_{S1} \cup A_{S2}$, $AG_{S}$ = $AG_{S1} \cup AG_{S2}$ and  $ \forall R_{iS} \in R_{s}$, $R_{iS} = R_{iS1} \cup R_{iS2}$; e.g. $ isTypeOf_{RS} = isTypeOf_{RS1}  \cup isTypeOf_{RS2} $ 
 \\}

\subsection{SOS Schemas}

As mentioned before, SOS allows the publication and consumption of information gathered by sensors or sensors system. Schemas associated to this service specification are probably among the most complex geospatial web service schemas, as they are built on the foundation provided by other specifications such as Geography Markup Language (GML) \cite{ogc:gml}, Sensor Model Language (SensorML) \cite{ogc:sml}, and Observation and Measurements (O\&M) \cite{ogc:om}. GML is a language for expressing geographical features, which is used as a common language through all of the OGC specifications. SensorML is a language used to describe sensors and sensor systems. And, O\&M is used as encoding for sensor observations. 

\begin{figure}[!h]
  \begin{center}
  \includegraphics[scale=0.7]{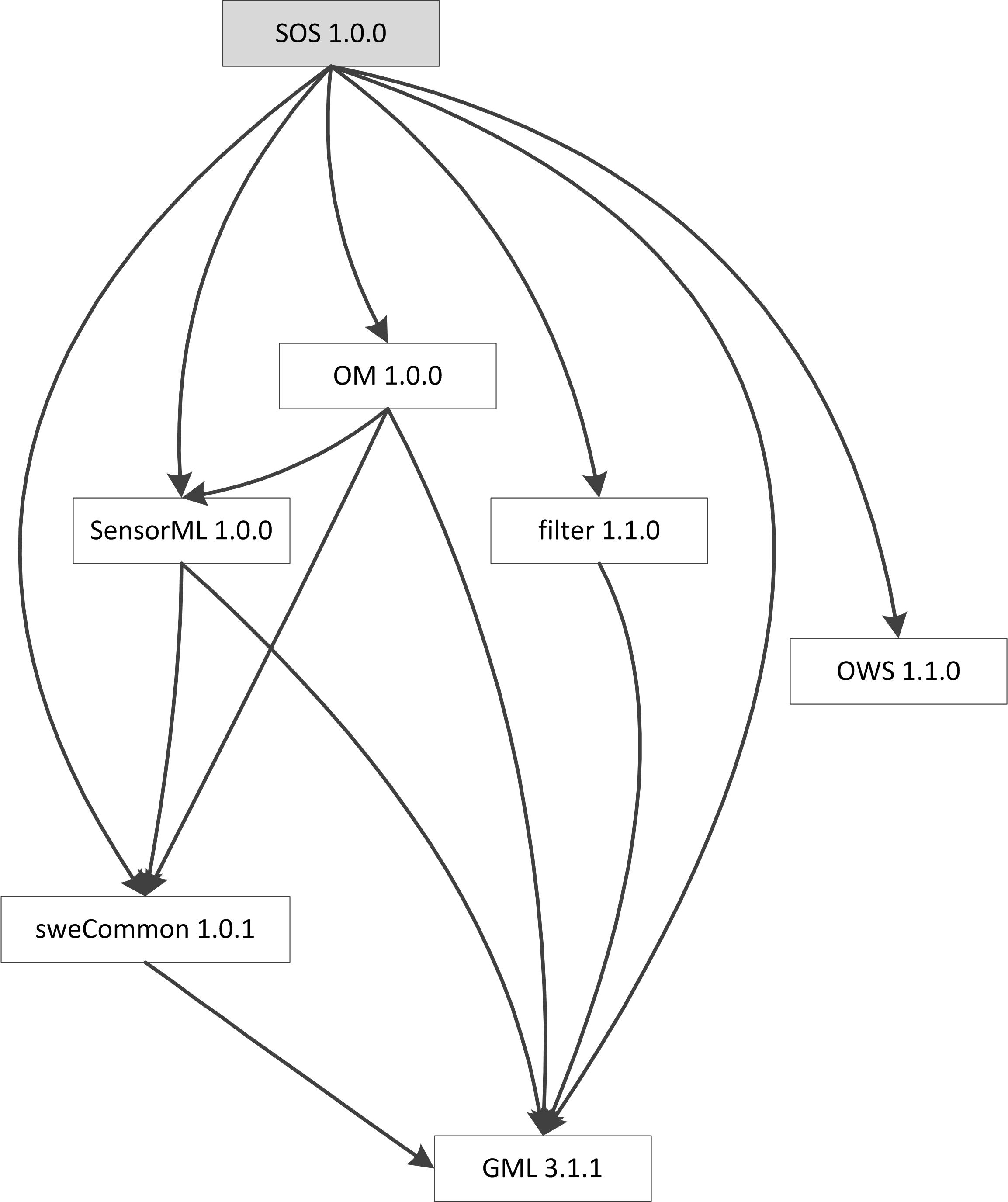}\\
  \caption{Dependencies of SOS schemas from other specifications}\label{fig6}
  \end{center}
\end{figure}

Figure \ref{fig6} shows the dependencies of schemas in SOS from schemas in other specifications. In addition to the specifications mentioned before, SOS depends also on OWS Commmon \cite{ogc:common}, for common mechanisms in OWS; Filter Encoding Implementation Specification \cite{ogc:filter}, to filter observations requested to the server; and  SWE Common\footnote{The first version of SWE Commmon is embedded on the SensorML implementation specification document, although its schemas are physically separated on a different folder}, that contains shared common data types and data encodings for all of the specifications related to sensors.

The whole SOS schema set contains more than 700 complex types and global elements, which make it large, according to the categorization for schema size based in the number of complex types presented in \cite{proc:lammel}. In this categorization a schema set with a number of complex types in the range 256-1,000 is considered large. Other categories are \textit{mini}, 0-32 complex types; \textit{small}, 32-100 complex types; \textit{medium}, 100 -256 complex types; and \textit{huge}, more than 1,000 complex types.

\section{Simplifying Schema Files}

In practical terms our problem of simplifying the schema set related to SOS, denoted as $S_{SOS}$, to the subset that is used in an actual implementation \textit{P}, denoted as $S_{P}$ could be formulated as follows:\\

\textit{\textbf{Problem}: Calculate $S_{P}$ starting from $S_{SOS}$ and \textit{X},  a set of instance files, knowing that $X \subseteq I(S_{P})$, trying to make $S_{P}$ as small as possible.}\\
 
As the set of valid instance files for a schema is potentially infinite, the resulting schema set should validate correctly all of the files in \textit{X} files, but might validate other instance files as well. 

The algorithm presented in this section is based on two main assumptions. The first one is that actual implementations do not use all of the information contained on the schemas. Instead, they use the only the parts required to implement specific application requirements. Although this assumption may seem obvious to some extent, it is supported by the results presented in \cite{proc:tamayo}, which showed that a set of 53 SOS server instances available on the Internet used less than 30\% of the SOS schemas. The second assumption is that a representative set of instance files is available beforehand to drive the simplification process. By representative we mean in this context that all of the information items on instance files that must be parsed by the application must be represented somehow in some of the input instance files. Unfortunately, if instances with new information must be added or existing instances are discarded the simplification algorithm must be executed again.

\subsection{Helper Functions}

The algorithm to calculate $S_{P}$ uses the following helper operations in its definition:

\begin{itemize}
\item \textit{\textbf{typeOf}(node)}: returns the type of an XML node in an instance file. For example in instance 1 in Figure 2, \textit{\textbf{typeOf}(}\textbf{/Container}\textit{) = ContainerType}, \textit{\textbf{typeOf}( }\textbf{/Container/item[1]}\textit{) = Base}.  In instance 2 \textit{typeOf( }\textbf{/Container/item[1]}\textit{) = Child}.
\begin{sloppypar}
\item \textit{\textbf{element}(node)}: returns the element definition matching the content of \textit{node}. For example, \textit{\textbf{element}(}\textbf{/Container/item[1]}\textit{) = ContainerType:item} in both instances in Figure 2.
\end{sloppypar}
\begin{sloppypar}
\item \textit{\textbf{containerOf}(node)}: returns the component containing the definition or reference to \textit{\textbf{element}(node)}. For example, \textit{\textbf{containerOf}(}\textbf{/Container/item[1]}\textit{) = \textbf{containerOf}(\textit{ContainerType:item}) = ContainerType}.
\end{sloppypar}
\item	\textit{\textbf{ancestors}(type)}: returns all of the ancestors of \textit{type}.
\begin{sloppypar}
\item \textit{\textbf{leaf}(node)} : returns \textit{true} is node is a leaf, i.e. \textit{node} does not contain any child element and has a value. Examples of leaf nodes in instance 2 in Figure \ref{fig2} are \textbf{/Container /item[1]/baseElem} and \textbf{/Container/item[1]/chdElem}. 
\end{sloppypar}
\item \textit{\textbf{root}(instance file)}: returns the root node of \textit{instance file}.
\begin{sloppypar}
\item \textit{\textbf{addValueToRelation}(S, R(x,y))}: adds \textit{R(x,y)} to the schema set \textit{S}. \textit{R} must be one of the relations defined in Section 2.1.
\end{sloppypar}
\item \textit{\textbf{copyRelations}($S_{T}$, $S_{S}$, $C$)}: Copy all relation pairs between schema components in $C$, from the source schema set $S_{S}$ to the target schema set $S_{T}$.
\end{itemize}

\subsection{Algorithm}
The algorithm to calculate $S_{P}$, henceforth called \textit{subsetting algorithm},  is expressed as follows:\\
\textit{
\begin{tabbing}
 \= Input: X = {x | x input instance file}\\
\> Input: schema set S\\
\> Output: schema subset  $S_{X}$ needed to validate instances in X\\
\> $S_{X} = \emptyset$,\\ 
\> For \= each x  in X   \\
\> \textbf{beginFor}\\
\> \> T = \textbf{SchemaSubsetUsedIn}(\textbf{root}(x), S)\\
\> \> $S_{X}$ = union($S_{X}$ , T)\\
\> \textbf{endFor}\\
\>Result = $S_{X}$\\
\end{tabbing}
}
\begin{sloppypar}
The key of this algorithm is the function \textit{\textbf{SchemaSubsetUsedIn}(node, schema set)} that calculates the subset of the schemas used in an XML file fragment starting at a given node. The second parameter is the schema set defining the fragment structure. The result of this function is calculated for the root element of all instance files and then joined through the \textit{union} operation defined in the previous section. 

Next, we present the algorithm for \textit{\textbf{SchemaSubsetUsedIn}}. For the sake of clarity in the exposition of the algorithm we do not consider attributes and substitution groups. The code considering these cases is similar to processing element and subtypes. \\
\end{sloppypar}

\begin{sloppypar}
\textit{
SchemaSubsetUsedIn
\begin{tabbing}
Input: instance file node x \\
Input: schema set S \\
Output: schema subset  $S_{x}$ needed to validate the \\ nodes contained in x\\
$S_{x} = \emptyset $ \\
$E_{Sx} = E_{Sx} + \textbf{element}(x)$ \\
$T_{Sx} = T_{Sx} + \textbf{typeOf}(x) + \textbf{ancestors}(\textbf{typeOf}(x))$\\
\textbf{addValueToRelation}(\= $S_{x}$,\textbf{typeOf} (\textbf{element}(x), \\ 
\> \textbf{typeOf}(\textbf{element}(x))))\\
\textbf{copyRelations}($S_{x}$, S, \textbf{ancestors}(\textbf{typeOf}(x)))\\
If  \= not \textbf{leaf}(x) Then\\
beginIf\\
\> For \= each child node z of x\\
\> beginFor\\
\> \> $S_{x}$ = union ($S_{x}$, \textbf{SchemaSubsetUsedIn}(z, S))\\
\> \> Container =\textbf{containerOf}(x) \\
\> \> If z \= belongs to a model group M Then \\
\>  \> beginIf \\
\>  \> \> $ MG_{Sx} =  MG_{Sx} $ + x;\\
 \> \> \> \textbf{addValue}\=\textbf{ToRelation}($S_{x}$, \\ 
  \> \> \>   \> \textbf{reference}(\textbf{containerOf}(M), M))\\
\> \> \> Container = M \\
\>  \> endIf  \\
\>  \> If z is reference to global element\\
\> \> \> \textbf{addValueToRelation}($S_{x}$, \\
  \> \> \>   \> \textbf{reference}(Container , \textbf{element}(z)))\\
\> \> Else \\
\> \> \> \textbf{addValueToRelation}($S_{x}$, \\
  \> \> \>   \> \textbf{contains}(Container , \textbf{element}(z)))\\
\> endFor\\
endIf\\
Result =$S_{x}$\\
\end{tabbing}
}
\end{sloppypar}

\begin{sloppypar}
The algorithm starts by adding  the element definition matching the content of the node specified as input to the result. The type of the node is also added, as well as pair \textit{(\textbf{element}(x), \textbf{typeOf}(\textbf{element}(x)))} to relation \textit{\textbf{isTypeOf}}. It is very important to notice at this point that \textit{\textbf{typeOf}(x)} and \textit{\textbf{typeOf}(\textbf{element}(x))} are not always the same because the dynamic type of \textit{x} may be a subtype of the declared type for the element matching its structure. This is the case in instance 2 listed in Figure \ref{fig2}, where \textit{\textbf{typeOf}(}\textbf{/Container/item[1]}\textit{)} = \textit{Child}, but \textit{\textbf{typeOf}(\textbf{element}(}\textbf{/Container/item[1]})\textit{)} = \textit{\textbf{typeOf}(ContainerType:item)} = \textit{Base}.  All of the ancestors of the type and all of their relations are also added to the result to maintain consistency of the model.

The next step is to analyse the child nodes in \textit{x}, in case it has any. For each child node \textit{z}, we call recursively the function \textit{\textbf{SchemaSubsetUsedIn}} and  the schema set returned by this function is combined with the current result using the \textit{union} operation. After this, a set of relation values are added to maintain the consistency of the model. First, the container of \textit{x} is calculated. This container is the type or model group that contains the element matching node \textit{x}. It could be \textit{\textbf{typeOf}(x)}, but could also be any of its ancestors. It also could be any model group referenced by \textit{\textbf{typeOf}(x)} or any of its ancestors. For example, let us calculate \textit{\textbf{containerOf}(}\textbf{/Container/item[1]/baseElem}\textit{)} in the second example in Figure \ref{fig2}. Even when \textit{\textbf{typeOf}(}\textbf{/Container/item[1]}\textit{)} is type \textit{Child}, this type does not contain the definition of \textbf{Container/item[1]/baseElem} because it was inherited from \textit{Base}. 

The relation between \textit{\textbf{element}(z)} and its container must be added to the result. The pair (\textit{\textbf{ContainerOf(element}(z))},\textit{\textbf{element}(z)}) is added to \textit{\textbf{reference}} or \textit{\textbf{contains}} depending if the element is referenced or it is a nested declaration. In the case the container is a model group the reference between its own container and the model group must be added to the result as well. 
\end{sloppypar}

\section{Experimentation}

In order to prove the effectiveness of the algorithm we have developed a prototype implementation for it.  After this we have used it with a real case study described in the next subsection.

\subsection{Case Study}
The case study is the implementation of the communication layer for a client for SOS targeted to the Android platform. The client must provide support for the Core Profile of the SOS specification, which includes the following operations \cite{ogc:sos}:

\begin{itemize}
\item \textit{GetCapabilities}: Operation to get metadata information about the service including title, keywords, provider information, supported operations, advertised observation offerings. 
\item\textit{DescribeSensor}: Operation to get information about a given sensor.
\item \textit{GetObservation}: Operation to get a set of observations from a given offering. The observations can be filtered by a time instant or interval, location, etc.
\end{itemize}

The common flow of interactions between SOS clients and servers starts when the client issues a \textit{GetCapabilities} request to the server, which answers by sending back its \textit{Capabilities} file. After parsing this file the client knows which operations are supported by the server and which information about sensors and observations can be requested by issuing \textit{DescribeSensor} and \textit{GetObservation} requests. 

\begin{figure}[!h]
  \begin{center}
  \includegraphics[scale=0.23]{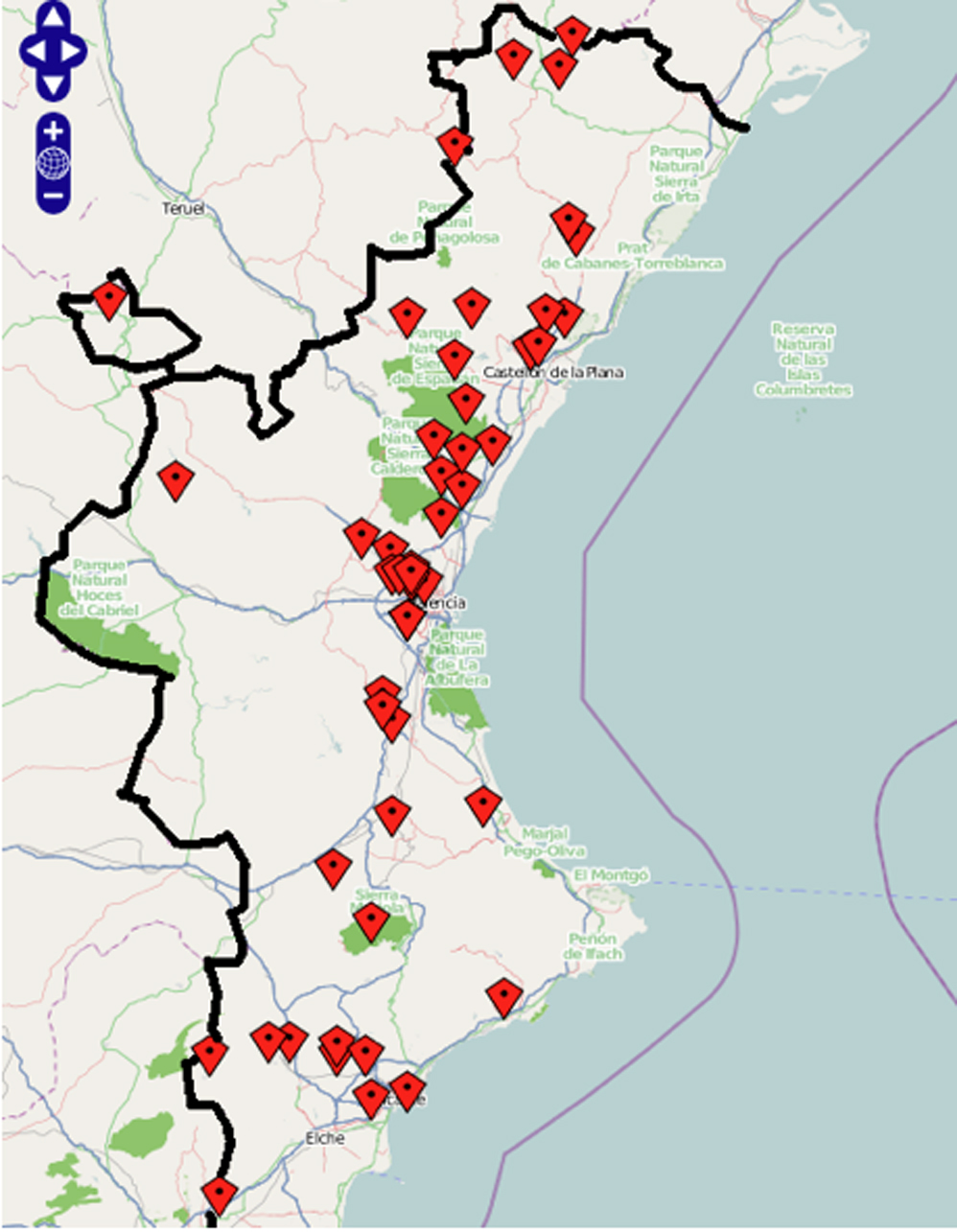}\\
  \caption{Location of air pollution control stations in the Valencian Community}\label{fig7}
  \end{center}
  
\end{figure} 

On the server side we use a 52$^{\circ}$ North SOS Server\footnote{http://52north.org/SensorWeb/sos/}, containing information about air quality for the Valencian Community gathered by 57 control stations located in that area (Figure \ref{fig7}). The stations measure the level of different contaminants in the atmosphere.

In order to measure how much the schemas can be reduced with the subsetting algorithm, we compare the size of the original or \textit{full schema set} with the size of the reduced or \textit{simplified schema set}. To measure the size of a schema set S = ($T_{S}$, $E_{S}$, $A_{S}$, $MG_{S}$, $AG_{S}$, $R_{S}$), we calculate the cardinality of the first five sets conforming the schema set, and the cardinalities of every relation included in  $R_{S}$.

\begin{figure}[!h]
  \begin{center}
  \includegraphics[scale=0.8]{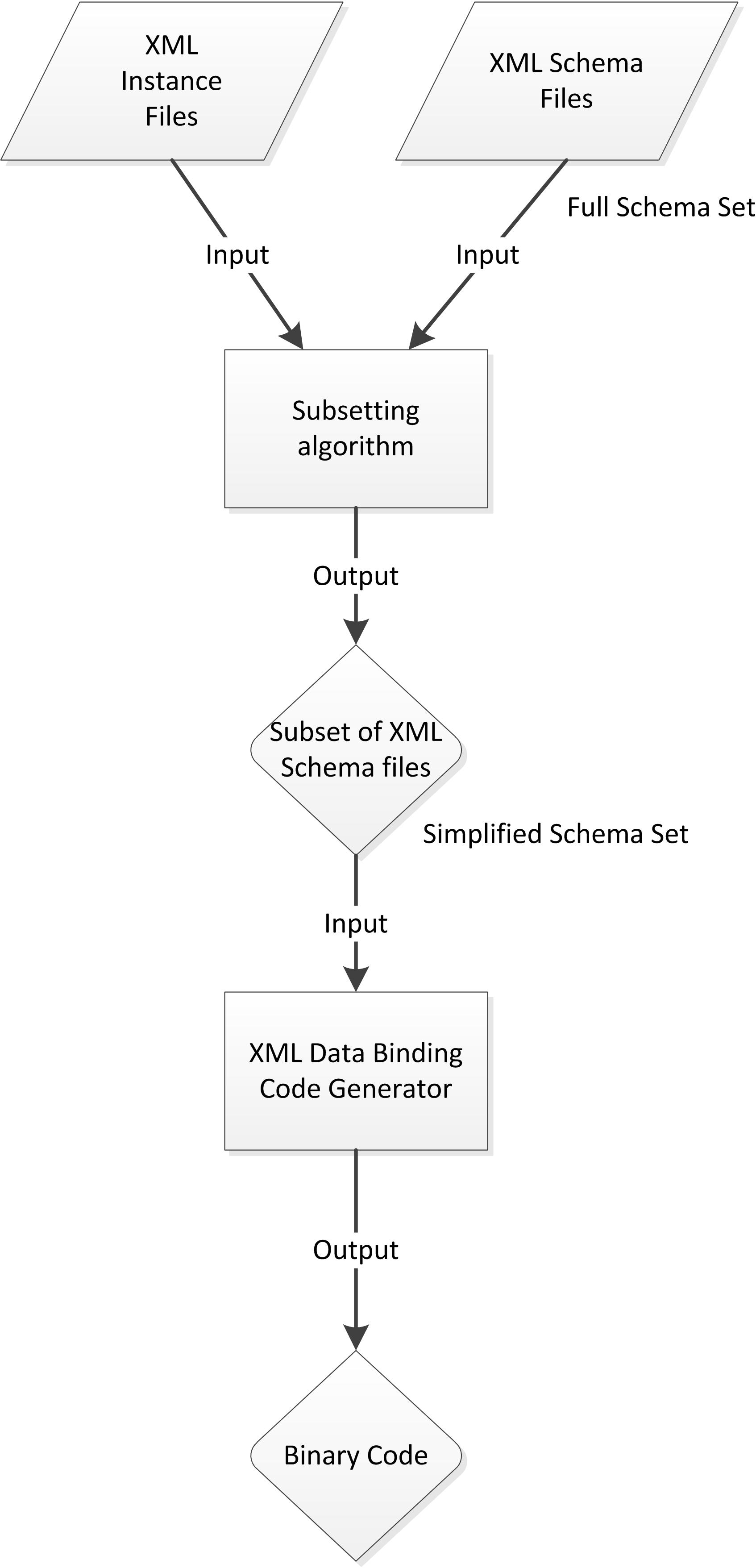}\\
  \caption{Flow diagram for the experiment}\label{flow}
  \end{center}
  
\end{figure}

After this, we use some code generators for XML data binding to measure how much the generated binary files are reduced when using the simplified schema set. The steps of the process followed during the experiment are shown in Figure \ref{flow}.

As our work is mostly focused on the production of XML processing code, we just consider this part of the implementation in the following subsections.

\subsection{Gathering Input Instance Files }

In order to generate the schema subset needed for the SOS client we must decide which input to pass to the algorithm. To obtain this set of instance files we sent requests manually to the server and stored server responses. We gathered 2492 instance files as input: the capabilities file, 2312 responses containing sensor descriptions, and 179 corresponding to observations. Our application must be capable of processing the following root elements:
\begin{sloppypar}
\begin{itemize}
\item \textit{Capabilities}: Server response with the service capabilities file
\item \textit{SensorML}: Server response containing information about a sensor.
\item  \textit{ObservationCollection}: Server response with observations data.
\end{itemize}
\end{sloppypar}
The first element is defined directly in the SOS specification and the other two are imported from the SensorML and O\&M specifications, respectively.  The number of files to be used as input will depend on the requirements of the particular application being developed. It might depend on availability of the instance files or on how different the content of these files is. In our case, although a considerably large number of input files was used, just a few would suffice because XML tags contained in sensor descriptions and observations files were basically the same within the two groups of files.

\subsection{Generating the Output Subset}

\begin{figure}[!h]
  \begin{center}
  \includegraphics[scale=0.115]{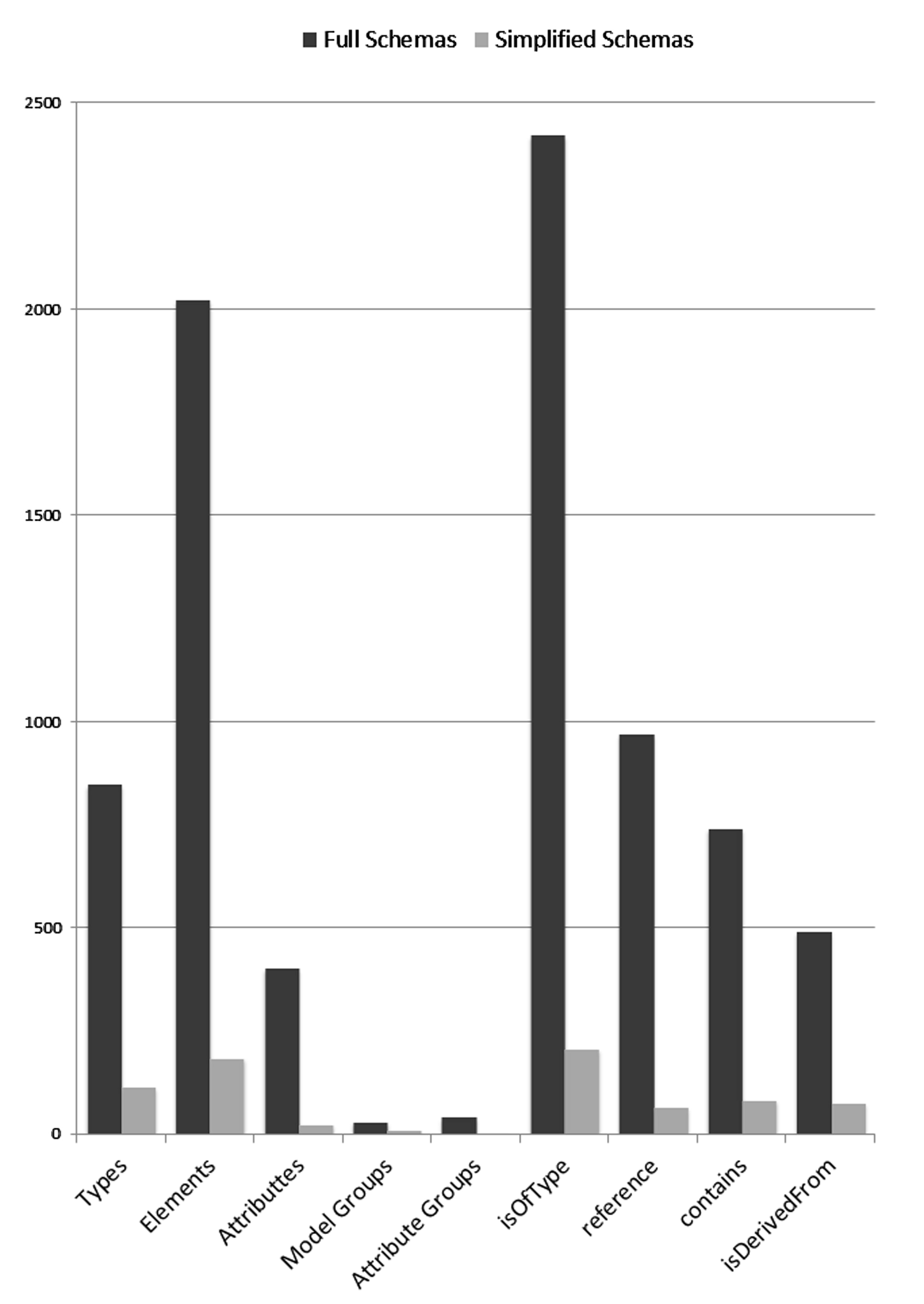}\\
  \caption{Simple schema metrics for original schemas and simplified schemas}\label{fig8}
  \end{center}
\end{figure}

After applying the algorithm with the input described above we obtained the results shown in Table \ref{tab1} and Figure \ref{fig8}, where the original schemas set is compared with the simplified set. In addition to cardinalities of components and relations we use two composite metrics: $Total_{C}$ for the summation of cardinalities of all components and $Total_{R}$ for the summation of cardinalities of relations. Results show that the subsetting algorithm allows a substantial reduction of the original schema set of about 90\% of its size. 

\begin{table}[!h]
\centering
\caption{Comparing original and simplified schema sets}\label{tab1}
\begin{tabular}{|p{3.8cm}|p{1.6cm}|p{1.6cm}|} \hline
Metric&Full Schema Set & Simplified Schema Set\\ \hline
$|T_{S}|$ &846&112\\\hline
$|E_{S}|$ &2020&183\\\hline
$|A_{S}|$ &400&22\\\hline
$|MG_{S}|$ &28&7\\\hline
$|AG_{S}|$ &39&3\\\hline
$|isTypeOf_{S}|$ &2420&205\\\hline
$|reference_{S}|$ &968&63\\\hline
$|contains_{S}|$ &739&81\\\hline
$|isDerivedFrom_{S}|$ &490&74 \\\hline
$|isInSubstitutionGroup_{S}|$ &290&17\\\hline
$|Total_{C}|$ &3333&327\\\hline
$|Total_{R}|$ &4617&423\\\hline
\end{tabular}
\end{table}

\subsection{Generating Binary Code}

We explore next how this reduction is translated into generated code, specifically we will use XBinder to generate code for the Android platform. XBinder is a XML data-binding generator that produces code for several programming languages (C, C++, Java, C\#). It also allows the generation of code targeted to different mobile platforms such as Android\footnote{http://www.android.com}  and CLDC \footnote{http://java.sun.com/products/cldc/}. We will use as well other generators targeted to the Java programming language, but not targeted to mobile devices: XMLBeans\footnote{http://xmlbeans.apache.org}  and JAXB-RI \footnote{https://jaxb.dev.java.net}, to show that our algorithm could be also useful to other kind of systems. 

The main metric used to compare generated code is \textit{size} measured in KiloBytes (KBs). Source code is generated for the schemas before and after the simplification algorithm is applied. Then, the source is compiled and compressed into a JAR file. All of the generators need, apart from the generated code, a set of supporting libraries, which is why we compared the size of the generated code with and without considering the supporting libraries. 

\begin{table}[!h]
\centering
\caption{Comparing size of generated code (KBs) for original and simplified schema sets}\label{tab2}
\begin{tabular}{|p{2cm}|p{1.5cm}|p{1.5cm}|p{1.5cm}|} \hline
 &XBinder & JAXB & XMLBeans\\ \hline
Full &3626&754&2822\\\hline
Reduced &567&90&972\\\hline
Libs &190&1,056&2,684\\\hline
Full+libs &3816&1810&8879\\\hline
Reduced+libs &684&1146&3655\\\hline
\end{tabular}
\end{table}

Table \ref{tab2} shows in the first two rows the comparison of the code size only for generated code (Full, Reduced) showing a large reduction of between 79 and 88\%. The metric values for the code generated using JAXB stand out among the rest because they are significantly smaller. The third row shows the size of the supporting libraries of each generator (Libs). At this point we can see the big difference in size between the XBinder libraries for Android and the JAXB and XMLBeans libraries. XBinder, when used to generate code for mobile devices, uses very light
supporting libraries, but at the expense of moving most of XML processing code to the generated code. On the other hand, JAXB has all of the XML processing code on the supporting libraries and generates clean and compact code. In the case of XMLBeans, judging for the large size of the generated code and supporting libraries it is clear that it has not been optimized to work with large schema sets.

Last two rows compare code size including supporting libraries (Full+libs, Reduced+libs). In this case, if we calculate the overall reduction in code size it will be smaller than the one presented before, as the size of the libraries remains constant. It ranges now from a 37\% reduction in JAXB to 84\% in XBinder. Nevertheless, in all cases the reduction of the generated code size is substantial. And the size of the code targeted to mobile devices (684 KB) seems like something that can be handled by modern devices. 

It is important to notice on this experiment that for none of generators the binary code could be directly produced without needing some kind of adjustments. The adjustments could
be related to manually modify the generated source code to avoid compilation errors, changing configuration parameters to avoid name clashes in GML related to case sensitivity or components with very similar names, or the failure of the generator to follow the intricate dependencies between schema components.

\section{Related Work}
As mentioned in the introductory section the solutions for achieving efficient processing of XML for mobile devices use compression techniques to reduce the size of XML-encoded \cite{proc:kabisch, proc:kangasharju2, w3c:exi}. These solutions requires that the server be aware of the compressed formats which cause that servers already online might not be accessible if they cannot be modified to support the aforementioned formats.

Regarding schemas transformation, the closer referent to the algorithm presented in this paper is the \textit{GML subsetting tool} \cite{ogc:gml}, which allows the extraction of GML schemas subsets called \textit{profiles}. This tool presents limitations such as it can only be applied to GML schemas, it does not handle polymorphic dependencies related to subtyping (Section 2).

Other products that can be compared with our algorithm are generators that make some attempt to simplify the final code structure, such as JiBX\footnote{http://jibx.sourceforge.net} or XML Schema Definition Tool\footnote{http://msdn.microsoft.com/ en-us/library/x6c1kb0s.aspx}  (henceforth called XSD.NET).  JiBX offers the option of restricting the generated code to only those parts of the schemas that are referenced from other schema components. Unfortunately, JiBX does not support dynamic typing of elements in instance files, preventing its use to process geospatial schemas. XSD.NET is provided as part of the development tools of the .NET Framework. XSD.NET is the only product known by the authors that performs optimizations while still preserving all of the type dependencies, \textit{explicit} and \textit{hidden} ones. Still, none of these tools allows information to be extracted from instance files to perform generated code customizations.

Regarding the use of instance files to drive the manipulation of schemas, a lot of work has been done related to \textit{schema inference}, where instance files are used to generated adequate schema files that can be used to assess their validity (e.g. \cite{proc:bex, proc:hegewald, proc:min, article:mlynkova}). This problem is different from the one presented here, where schemas already exists, but must be refined to adjust to more specific requirements.

\section{Conclusions}
In this paper we have presented an algorithm to generate efficient code for XML data binding for mobile SOS-based applications. The algorithm take advantage of the fact that individual implementations use only portions of the standards' schemas allowing particular customizations to be applied by simplifying large schema sets, such the one associated to SOS, in an application-specific manner by using a set of XML instance files conforming to these schemas. 

Results of applying the algorithm to a real-world use case scenario have shown that the algorithm allows a substantial reduction of the original schema set of about the 90\% of its size. This huge reduction in schema size is translated into a reduction of generated binary code of more than 80\% of its size for a SOS client targeted to the Android platform. As the transformation is done at the schema level and no assumption about the target platform is made by the algorithm it still can be used for other kind of SOS applications. Nevertheless, the resource constraints associated to mobile devices make the algorithm far more useful in this area. 

This algorithm could be also applied to other OWS specifications although based on the little experience of authors with other specifications besides SOS, we cannot state that the reduction could be as large as that obtained in the use case scenario presented in Section 4. 

Further work will integrate the algorithm to a code generator targeted to mobile devices. This code generator will take advantage of other useful information that can be extracted during the simplification process that will allow to optimize further the generated code. In addition, a performance study for the generated code would be valuable, including also other aspects  such as memory consumption or execution speed.

\section{Acknowledgments}
This work has been partially supported by the "Espa\~{n}a Virtual" project (ref. CENIT 2008-1030) through the Instituto Geogr\'{a}fico Nacional (IGN). 
%
\bibliographystyle{abbrv}
\bibliography{sigproc}  

\end{document}